\documentstyle[12pt]{article}
\begin{document}
\begin{flushright}
hep-th/0302224\\
SNBNCBS-2003
\end{flushright}
\vskip 2.5cm
\begin{center}
{\bf \large { HAMILTONIAN AND LAGRANGIAN \\ DYNAMICS 
IN A NONCOMMUTATIVE SPACE}}

\vskip 2.5cm

{\bf R.P.MALIK}
\footnote{ E-mail address: malik@boson.bose.res.in  }\\
{\it S. N. Bose National Centre for Basic Sciences,} \\
{\it Block-JD, Sector-III, Salt Lake, Calcutta-700 098, India} \\

\vskip 2.5cm

\end{center}

\noindent
{\bf Abstract}:
We discuss the dynamics of a particular two-dimensional (2D) physical system 
in the four dimensional (4D) (non-)commutative phase space by exploiting the 
consistent Hamiltonian and Lagrangian formalisms based on the symplectic 
structures defined on the 4D (non-)commutative cotangent manifolds. The 
noncommutativity exists {\it equivalently} in the coordinate or the momentum 
planes embedded in the 4D cotangent manifolds. The signature of
this noncommutativity is 
reflected in the derivation of the first-order Lagrangians where we exploit 
the most general form of the Legendre transformation defined on the 
(non-)commutative (co-)tangent manifolds. The second-order Lagrangian, defined 
on the 4D {\it tangent manifold}, turns out to be  the {\it same} irrespective 
of the noncommutativity present in the 4D cotangent manifolds for the 
discussion of the Hamiltonian formulation.
A connection with the noncommutativity of the dynamics, associated with the 
quantum groups on the $q$-deformed 4D cotangent manifolds, is also pointed out.
\\
 
$~~~~~~~~~~~~~~~~$manifolds; symplectic structures; 
quantum groups\\


\baselineskip=16pt


\newpage

\noindent
{\bf 1 Introduction}\\

\noindent
In recent years, there has been a great deal of interest in the study
of noncommutative geometry because of its 
very neat and natural appearance in the context of
brane configurations related to the dynamics of string theory and 
matrix model of the M-theory [1-3]. It has been demonstrated that, in a certain
specific limit, the whole string dynamics can be described by the minimally
coupled gauge theory on a noncommutative space [2]. Even the experimental 
test for the existence of such kind of noncommutativity has been 
proposed where the synchrotron radiation has been incorporated in the 
experimental set up in a very specific way [4]. In such kinds of 
experimental proposals,
it has been argued that only the quantum mechanical approximations are good
enough to shed some light on the existence of the noncommutativity [5,6]. This 
is why a whole lot of studies have gone into the understanding of 
quantum mechanics of some physical systems on the noncommutative spaces 
(see, e.g., [7-10] and references therein for details). In particular,
the motion of the charged particles in the 2D plane, under the influence of 
a constant perpendicular magnetic field, has been studied
extensively because of the fact that the noncommutativity appears 
in this problem (i.e. Landau problem) quite naturally [7-10].

In the present scenario of the frontier research in the realm of high energy
physics (in particular, 
connected with the string theory), the noncommutative (NC) effects are
presumed to show up {\it only} at very high energy scale 
(perhaps, very near to the Planck scale). One can observe, however, the
physical consequences of these NC effects in the low energy effective actions
for the physically interesting systems.
Equivalently, on the other hand, one can construct the low energy effective 
actions for the above physical systems by exploiting the basic ideas 
behind the NC geometry that are supposed to be the benchmarks of physics
at very high energy scale. In the latter category, mention can be made of a
couple of interesting examples such as the NC Chern-Simons theory [11] 
and the NC extension of the standard model [12].
One of the most interesting physical system that encompasses the NC structure
in its folds at low energy scale is the 2D Landau problem.
The purpose of our present paper is to study, systematically, the Hamiltonian
and Lagrangian formulation of the 2D Landau problem based on the symplectic
structures defined on the 4D cotangent manifolds. We lay emphasis on 
{\it the classical equations of motion} for the charged particle moving
on a 2D plane under the influence of a harmonic oscillator potential and
a constant magnetic field perpendicular to this 2D plane. We demonstrate that
there exist three consistent
first-order Lagrangians (FOLs) and three Hamiltonians that lead
to the same classical equations of motion when we exploit 
the Euler-Lagrange equations of motion and the Hamilton's equations of motion
using (i) the canonical Poisson brackets (PBs), and 
(ii) the nontrivial (noncommutative) Poisson brackets. The nontrivial
noncommutativity is found to be {\it equivalently}
present either in the coordinate plane or in 
the momentum plane, embedded in the 4D cotangent manifold.
The FOLs are defined 
and derived by the most general form of the Legendre transformations 
which exploit the above symplectic structures (see, e.g., [13] for details). 
It is interesting to point out the fact that the second-order Lagrangian
$L^{(s)} (x, \dot x)$, defined in the 4D tangent manifold, turns out to
be the same for all the three consistent
FOLs and the Hamiltonians. In fact, the signature
of the NC structure, present in the PBs, disappears in the derivation
of the second-order Lagrangian (SOL) for the system under consideration. This
happens, perhaps, because of the fact that the Hamiltonians (and the
corresponding PBs) are connected with one-another due to a {\it special class}
of transformations (see, e.g., (3.5) and (3.6) below)
on the phase variables under
which {\it the sum of the areas of the phase space remains unchanged}.
Furthermore, we have shown  that these special 
class of transformations are {\it not exactly} the canonical transformations
because the form of the {\it basic} canonical PBs does not remain form
invariant. In particular, the brackets {\it equivalently}
defined in the 2D coordinate or momentum plane
{\it do change} their form (under the above transformations) which is
the root cause of the presence of NC structures in the theory. In the framework
of quantum groups, we demonstrate and establish that the above (phase space)
area preserving transformations correspond to the $SL_{q, q^{-1}} (2)$ 
invariant transformations on the phase variables defined on the 4D 
$q$-deformed cotangent manifolds. The consistency of the above
transformations with the structure of the $q$-deformed
PBs, however, puts a restriction $q^2 = 1$ on the deformation parameter.
This requirement, in turn, allows only the 
presence of exact canonical transformations
for the phase variables in the framework of transformations
generated by the quantum groups.

The contents of our present paper are organized as follows. In section 2,
we discuss the canonical brackets and show that the classical dynamical
equations of motion for the charged particles on the 2D plane
(moving under the influence of a constant perpendicular magnetic field)
can be derived from a canonical Hamiltonian. Section 3 is devoted to the
discussion of the nontrivial NC brackets which are exploited in the
derivation of the same classical dynamical equations from another set of
a couple of Hamiltonians. In this section, we also discuss about the
special kind of transformations on the phase space variables that connect
the (non-)trivial PBs as well as the Hamiltonian functions. We deal
with the noncommutativity associated with the quantum groups 
in section 4 and show its
connection with the NC structures associated with the PBs, discussed in the 
previous sections. Finally, in section 5, we concisely summarize our results 
and make some concluding remarks.\\ 

\noindent
{\bf 2 Canonical brackets: Hamiltonian and Lagrangian dynamics}\\

\noindent
We start off with the following Hamiltonian, describing the motion
of the charged particles in a 2D plane
under the influence of a two-dimensional
harmonic oscillator potential, as
\footnote{ We adopt here the summation convention  where all the repeated
Latin indices $i, j, k.....= 1, 2$ in the coordinate and/or momentum space
and $A, B, C.....= 1, 2, 3, 4$ in the momentum phase 
(i.e. cotangent) space are
summed over. The antisymmetric Levi-Civita tensor $\varepsilon_{ij}$ is
chosen such that $\varepsilon_{12} = + 1 = - \varepsilon^{12}$.}
$$
\begin{array}{lcl}
H_{1} = {\displaystyle \frac{1}{2}\; p_{i} \;p_{i} + \frac{1}{2}\; 
\Bigl (\omega^2  + \frac{k^2}{4} \Bigr )\;
x_{i}\; x_{i} - \frac{k}{2}\;  \varepsilon_{ij}\; x_{i} \;p_{j}},
\end{array} \eqno(2.1)
$$
where $(x_{i}, p_{i})$ (with $ i = 1, 2$) are the canonically conjugate
position and momenta variables for the charged particles with individual 
charge $e$ and $\omega$ is the frequency of the harmonic
oscillations. The constant $k = \frac{e}{c} {\cal B}$ is connected with 
the charge $e$, the speed of light $c$ and the magnetic field 
${\cal B}$ perpendicular to the 2D plane of oscillation. Here the
mass of the particle has been set equal to one for the sake of simplicity
of all the later calculations. The dynamical equation of motion for this 
physical system
$$
\begin{array}{lcl}
\ddot x_{i} = - \omega^2 \; x_{i} + k\; \varepsilon_{ij}\; \dot x_{j},
\end{array} \eqno(2.2)
$$
can be derived from the above Hamiltonian by exploiting the following
canonical Poisson brackets (PBs) among the conjugate variables
$$
\begin{array}{lcl}
\Bigl \{ x_{i}, p_{j} \Bigr \}_{(PB)}^{(x,p)} = \delta_{ij},\; \qquad
\Bigl \{ x_{i}, x_{j} \Bigr \}_{(PB)}^{(x,p)} = 0,\; \qquad 
\Bigl \{ p_{i}, p_{j} \Bigr \}_{(PB)}^{(x,p)} = 0. 
\end{array} \eqno(2.3)
$$
With the help of the above canonical brackets (2.3), the canonical 
symplectic structures on the 4D cotangent manifold, for 
the Hamiltonian (2.1), can be defined as
$$
\begin{array}{lcl}
\Omega^{AB} (1) =
\left (\begin{array} {cccc}
0 & 0 & 1 & 0 \\
0 & 0 & 0 & 1 \\
- 1 & 0 & 0 & 0\\
0 & - 1 & 0 & 0\\
\end{array} \right ),
\qquad
\Omega_{AB} (1) =
\left (\begin{array} {cccc}
0 & 0 & - 1 & 0 \\
0 & 0 & 0 & - 1 \\
1 & 0 & 0 & 0\\
0 & 1 & 0 & 0\\
\end{array} \right ),
\end{array} \eqno(2.4)
$$
where the entry $(1)$ in the parenthesis of the symplectic structures
$\Omega^{AB}$ and $\Omega_{AB}$ stand for the subscript of the
Hamiltonian in (2.1) and the
notation $z^A = (z^1 , z^2, z^3, z^4) \equiv (x_1, x_{2},
p_1, p_2)$ has been introduced in the definition of the 
{\it matrix form} of the symplectic structure as given below
$$
\begin{array}{lcl}
\Omega^{AB} (1)
= {\mbox{Matrix}}\; \Bigl ( \bigl \{ z^A, z^B \bigr \}_{(PB)} \Bigr ),\;
\qquad \Omega^{AB}\; \Omega_{BC} = \delta^A_C 
= \Omega_{CM} \;\Omega^{MA}.
\end{array} \eqno(2.5)
$$
The most general form of the PB between two dynamical variables 
$ f (z), g (z)$ in the momentum phase (i.e. cotangent) space can be defined as
$$
\begin{array}{lcl}
\Bigl \{ f (z),  g (z) \Bigr \}_{(PB)} = \Omega^{AB}\;
\partial_A f (z)\; \partial_{B} g (z),\; \qquad
\partial_A = {\displaystyle \frac{\partial} {\partial z^A}}.
\end{array} \eqno(2.6)
$$
In general, the symplectic structures in (2.4) and/or (2.5) can be 
functions of the phase variables $z^A$. In such a situation, the most
general form of the Legendre transformation (which leads to the
derivation of the first-order Lagrangian $L^{(f)}$) is given by [13]
$$
\begin{array}{lcl}
L^{(f)} (z, \dot z) = {\displaystyle z^A\; \bar \Omega_{AB} (z) \;\dot z^B 
- \;H (z)},
\end{array} \eqno(2.7)
$$
where the general form of the covariant symplectic
structure $\bar \Omega (z)$ is [13]
$$
\begin{array}{lcl}
\bar \Omega (z) = {\displaystyle \int_{0}^{1}}\; d \alpha \;\alpha\; 
\Omega (\alpha z).
\end{array} \eqno(2.8)
$$
For our case of canonical symplectic structures (listed in (2.4)
and satisfying (2.5)), the above formula
yields $ \bar \Omega_{AB} (1) = \frac{1}{2}\; 
\Omega_{AB} (1)$. The first-order Lagrangian (modulo some total 
derivatives w.r.t. time), for our discussion, is
$$
\begin{array}{lcl}
L_{1}^{(f)} (z, \dot z) &=& {\displaystyle 
\frac{1}{2}\; z^A \;\Omega_{AB} (1) \;\dot z^B - \;H_{1} (z)},\nonumber\\
&\equiv& p_{i} \;\dot x_{i}
- {\displaystyle \frac{1}{2}\; p_{i} \;p_{i} - \frac{1}{2} \;\Bigl (\omega^2 
+ \frac{1}{4} k^2 \Bigr )\;
x_{i}\; x_{i} + \frac{1}{2}\; k \;\varepsilon_{ij} \;x_{i} \;p_{j}.}
\end{array} \eqno(2.9)
$$
The equations of motion, emerging from the above FOL (2.9), are 
$$
\begin{array}{lcl}
\dot x_i = {\displaystyle 
p_i + \frac{k}{2}\; \varepsilon_{ij}\; x_j, \; \qquad
\dot p_i = - \Bigl (\omega^2 + \frac{1}{4}\; k^2 \Bigr )\; x_i 
+ \frac{1}{2}\; k\; \varepsilon_{ij}\; p_j.}
\end{array}\eqno(2.10)
$$
Combined together, the equations of motion (2.10)
imply our starting
equation of motion (2.2) that was derived from the Hamiltonian
(2.1) by exploiting the canonical PBs of (2.3).

Now, inserting  the equations of motions derived
from FOL (2.9) into itself, we obtain the following
second-order Lagrangian 
$$
\begin{array}{lcl}
L^{(s)} (x, \dot x)= {\displaystyle
\frac{1}{2}\; \dot x_i\; \dot x_{i} + \frac{k}{2}\; \varepsilon_{ij}
\;x_i\;\dot x_j - \frac{1}{2} \;\omega^2\; x_i\; x_i.}
\end{array}\eqno(2.11)
$$
A couple of comments are in order which would turn out to be quite handy for
our further discussions in the later sections. 
First and foremost, the Euler-Lagrange equation
of motion, emerging from (2.11), is nothing but our starting equation of 
motion (2.2). 
Second, the canonical momentum
$p_i$ defined from the above SOL in (2.11)
$$
\begin{array}{lcl}
p_i = {\displaystyle \frac{\partial L^{(s)} (x,\dot x)}{\partial \dot x_i}}
= \dot x_i - {\displaystyle \frac{k}{2}} \varepsilon_{ij}\; x_j,
\end{array} \eqno(2.12)
$$
turns out to be the same as the one derived from the Hamilton's
equations of motion by exploiting the Hamiltonian function (2.1). \\

\noindent
{\bf 3 Nontrivial brackets: Hamiltonian and Lagrangian dynamics}\\

\noindent
Keeping the dynamical evolution 
($\ddot x_i = -\; \omega^2\; x_i + k \;\varepsilon_{ij}\; \dot x_j$)
of the system intact, it is straightforward
to check that the following new Hamiltonians 
$$
\begin{array}{lcl}
H_{2} = {\displaystyle \frac{1}{2}\; P_{i} \;P_{i} + \frac{1}{2} \;\omega^2\; 
x_{i}\; x_{i}},
\end{array} \eqno(3.1)
$$
$$
\begin{array}{lcl}
H_{3} = {\displaystyle \frac{1}{2}\; p_{i} \;p_{i} + \frac{1}{2} \;\omega^2\; 
X_{i}\; X_{i}},
\end{array} \eqno(3.2)
$$
expressed in terms of the phase variables $(x_{i}, P_i)$ and
($X_i, p_i$) lead to our starting equation of motion (2.2) if we exploit the 
following {\it nontrivial} PBs in the phase space
$$
\begin{array}{lcl}
\Bigl \{ x_{i}, x_{j} \Bigr \}_{(PB)}^{(x,p)} = 0,\; \qquad
\Bigl \{ x_{i}, P_{j} \Bigr \}_{(PB)}^{(x,p)}  = \delta_{ij},\; \qquad
\Bigl \{ P_{i}, P_{j} \Bigr \}_{(PB)}^{(x,p)} = k \; \varepsilon_{ij}, 
\end{array} \eqno(3.3)
$$
$$
\begin{array}{lcl}
\Bigl \{ X_{i}, X_{j} \Bigr \}_{(PB)}^{(x,p)} 
= {\displaystyle \frac{k}{\omega^2}\;\varepsilon_{ij}}, \;\qquad
\Bigl \{ X_{i}, p_{j} \Bigr \}_{(PB)}^{(x,p)} = \delta_{ij},\; \qquad
\Bigl \{ p_{i}, p_{j} \Bigr \}_{(PB)}^{(x,p)} = 0. 
\end{array} \eqno(3.4)
$$
Here the new phase variables $(x_i, P_i)$ and $(X_i, p_i)$ parametrize
the transformed versions of the cotangent manifolds that are obtained by
exploiting a certain specific 
types of transformation on the original phase variables
$(x_i, p_i)$ (which earlier described the original cotangent manifold).
Furthermore, it is evident that the brackets in (3.3) and (3.4) are
defined for the Hamiltonians in (3.1) and (3.2), respectively.

As far as the the {\it nontrivial} PBs in (3.3) and (3.4) are concerned,
the noteworthy points, at this stage, are as follows. First, it is
interesting to pin-point that the PB between the co-ordinates
and momenta variables remain {\it the same} in the canonical brackets (2.3) 
as well as the nontrivial brackets (3.3) and (3.4) 
even though, as is clear, a certain specific
types of transformation have been performed on the
original phase space variables $(x_i, p_i)$ to make them the new
pairs of variables $(x_i, P_i)$ and $(X_i, p_i)$, respectively. 
Second, for the Hamiltonian (3.1),
it is the momenta variables $P_i$ (describing a momentum plane in the
transformed cotangent manifold) that are required to be noncommutative 
({\it vis-{\` a}-vis} the brackets in (2.3)) for the dynamical equations of 
motion (2.2) to remain intact. Third, it is the coordinates $X_i$ (defining 
a coordinate plane in the transformed cotangent manifold) that are required to
possess noncommutative nature ({\it vis-{\` a}-vis} the brackets in (2.3)) 
for the starting equation of motion (2.2) to remain unchanged. Fourth, it 
can be checked that the following specific transformations on the phase
variables
$$
\begin{array}{lcl}
p_{i} \rightarrow P_{i} = p_{i} + 
{\displaystyle \frac{k}{2}}\; \varepsilon_{ij}\; x_{j},\;
\qquad x_{i} \rightarrow X_{i} = x_{i},\; 
\qquad H_{1} (z) \rightarrow H_{1}^\prime (z) = H_{2} (z),
\end{array} \eqno(3.5)
$$
$$
\begin{array}{lcl}
p_{i} \rightarrow P_{i} = p_{i},\; \qquad
x_{i} \rightarrow X_{i} = x_{i} - 
{\displaystyle \frac{k}{2 \omega^2}}\; \varepsilon_{ij} \;p_{j},\;
\qquad \tilde H_{1} (z) \rightarrow \tilde H_{1}^\prime (z) = H_{3} (z),
\end{array} \eqno(3.6)
$$
relate $H_1 (z)$ to $H_2 (z) $ and $\tilde H_1 (z)$ to $H_3 (z)$ where (i) 
the phase variables $z$'s come in different guises for the different types
of Hamiltonian, and (ii) the Hamiltonian $\tilde H_{1} (z)$
$$
\begin{array}{lcl}
\tilde H_{1} (z) = {\displaystyle
\frac{1}{2}\; \Bigl ( 1 + \frac{1}{4}\; \frac{k^2}{\omega^2}
\Bigr )\;
p_{i}\; p_{i} + \frac{1}{2}\; \omega^2 \; x_{i} \;x_{i}
- \frac{k}{2}\; \varepsilon_{ij}\; x_{i}\; p_{j}},
\end{array} \eqno(3.7)
$$
is obtained from $ H_{1} (z)$ by the following canonical transformations
\footnote{ A comment on this by an anonymous theoretical physicist
is gratefully acknowledged.}
$$
\begin{array}{lcl}
&&x_{i} \rightarrow X_i = \omega^{-1}\; p_i,\; \qquad 
p_i \rightarrow P_i = - \omega\; x_{i},\; \nonumber\\
&& \Bigl \{ X_{i}, P_{j} \Bigr \}_{(PB)}^{(x,p)} = \delta_{ij},\; \qquad
\Bigl \{ X_{i}, X_{j} \Bigr \}_{(PB)}^{(x,p)} =
\Bigl \{ P_{i}, P_{j} \Bigr \}_{(PB)}^{(x,p)} = 0,
\end{array} \eqno(3.8)
$$
which show that $H_1 (z)$ and $\tilde H_1 (z)$ are 
canonically equivalent. Fifth, the transformations in (3.5) and (3.6)
imply the noncommutativity present in (3.3) and (3.4). Sixth, the
canonical equivalence of $H_1 (z)$ and $\tilde H_1 (z)$ demonstrates
that the noncommutativity of (3.3) and (3.4) are equivalent too.
Seventh, it is obvious that,
under the above transformations, the sum of the area elements in the
momentum phase (cotangent) space remain invariant (i.e. $ d X_i \; d P_i 
= d x_i\; d p_i $) for the transformations in (3.5) and (3.6).

With the nontrivial brackets defined in (3.3), it can be seen that the
analogues of the symplectic structures in (2.4) (and satisfying (2.5))
for the transformed cotangent manifold are 
$$
\begin{array}{lcl}
\Omega^{AB} (2) =
\left (\begin{array} {cccc}
0 & 0 & 1 & 0 \\
0 & 0 & 0 & 1 \\
- 1 & 0 & 0 & k\\
0 & - 1 & - k & 0\\
\end{array} \right ),
\qquad\;
\Omega_{AB} (2) =
\left (\begin{array} {cccc}
0 & k & - 1 & 0 \\
- k & 0 & 0 & - 1 \\
1 & 0 & 0 & 0\\
0 & 1 & 0 & 0\\
\end{array} \right ),
\end{array} \eqno(3.9)
$$
where the notation $z^A = (z^1, z^2, z^3, z^4) = (x_1, x_2, P_1, P_2)$
and the definitions of (2.5) have been taken into account. In terms of the
covariant symplectic structure $\Omega_{AB} (2)$, defined on the transformed
cotangent manifold, we obtain the following FOL (modulo some total 
derivatives w.r.t. time) for the case under consideration; namely,
$$
\begin{array}{lcl}
L_{2}^{(f)} (z, \dot z) &=& {\displaystyle 
\frac{1}{2}\; z^A \;\Omega_{AB} (2) \;\dot z^B - \;H_{2} (z) },\nonumber\\
&\equiv& P_{i}\; \dot x_{i}
+ {\displaystyle \frac{1}{2}}\; k \;\varepsilon_{ij}\; x_{i}\; \dot x_{j}
- {\displaystyle
\frac{1}{2}\;
P_{i}\; P_{i} - \frac{1}{2} \;\omega^2 \; x_{i} \;x_{i}.}
\end{array} \eqno(3.10)
$$
The following equations of motion emerge from the above FOL
$$
\begin{array}{lcl}
P_i = \dot x_i,\; \qquad \dot P_i = - \omega^2 \; x_i + k\;\varepsilon_{ij}\;
\dot x_j,
\end{array}\eqno(3.11)
$$
which, ultimately, lead to the derivation of the dynamical equation of motion
(2.2), we started with. It will be noticed that the above equations of motion 
(3.11) are quite different in appearance from their counterparts in (2.10)
which have been derived from the FOL $L^{f}_{(1)}$.
Now, exploiting (3.11), it is straightforward to check that the SOL
(which emerges from the FOL in (3.10)) is {\it exactly} the same as (2.11).

Now let us concentrate on the nontrivial brackets defined in (3.4). Here
the analogues of the symplectic structures in (2.4) (that satisfy (2.5)) are
$$
\begin{array}{lcl}
\Omega^{AB} (3) =
\left (\begin{array} {cccc}
0 & (k/\omega^2) & 1 & 0 \\
-  (k/\omega^2) & 0 & 0 & 1 \\
- 1 & 0 & 0 & 0\\
0 & - 1 & 0 & 0\\
\end{array} \right ),
\;\;
\Omega_{AB} (3) =
\left (\begin{array} {cccc}
0 & 0 & - 1 & 0 \\
0 & 0 & 0 & - 1 \\
1 & 0 & 0 & (k/\omega^2)\\
0 & 1 & - (k/\omega^2) & 0\\
\end{array} \right ),
\end{array} \eqno(3.12)
$$
where the notation $z^A = (z^1, z^2, z^3, z^4) = (X_1, X_2, p_1, p_2)$
has been chosen in the analogues of the definitions (2.5) for the
case under consideration. With the help of the covariant symplectic
structure in (3.12), one can obtain the FOL $L_{3}^{(f)}$ (modulo some
total derivatives w.r.t. time) as 
$$
\begin{array}{lcl}
L_{3}^{(f)} (z, \dot z) &=& {\displaystyle 
\frac{1}{2}\; z^A \;\Omega_{AB} (3)\; \dot z^B - \;H_{3} (z) },\nonumber\\
&\equiv& p_{i}\; \dot X_{i}
+ {\displaystyle
\frac{k}{2 \omega^2}\;  \varepsilon_{ij}\; p_{i} \dot p_{j}}
- {\displaystyle
\frac{1}{2}\;
p_{i}\; p_{i} - \frac{1}{2} \omega^2 \; X_{i} \;X_{i}.}
\end{array} \eqno(3.13)
$$
The above Lagrangian leads to the derivation of the Euler-Lagrange
equations of motion that are first-order in time derivative on 
$X_i$ and $p_i$. These $\dot X_i$ and 
$\dot p_i$ are intertwined together in one equation as
$$
\begin{array}{lcl}
\dot X_i = p_i - {\displaystyle \frac{k}{\omega^2}}\; \varepsilon_{ij}\;
\dot p_j,\; \qquad \;\dot p_i = -\; \omega^2\; X_i.
\end{array}\eqno(3.14)
$$
Combined together, the above equations of motion
lead to the derivation of the same dynamical equations of motion
(i.e. $\ddot X_i = - \omega^2 \; X_i + k\; \varepsilon_{ij}\; \dot X_j$)
as we started with in (2.2). It will be noticed that even though the equations
of motion are expressed, ultimately, in $X_i, \dot X_i, \ddot X_i$ (which are
 different from $x_i, \dot x_i, \ddot x_i$
in (2.2)), the evolution of the dynamics of the 
physical system w.r.t. time is the same as in (2.2).
It is very interesting to check that the SOL (that emerges from the FOL (3.13))
has {\it exactly} the same appearance as the one in (2.11), albeit 
expressed in terms of $X_i$ and $\dot X_i$. 

A few comments are in order as far as the transformations (3.5) and (3.6)
of the phase variables (defining the transformed cotangent space) are concerned.
It should be noticed that, unlike transformations (3.8),
these transformations are not {\it exactly}
canonical in nature even though they preserve the invariance of the sum of
the area elements in the cotangent manifold. To be precise, it should
be noted that, for a transformation $(x_{i}, p_{i}) \rightarrow (X_i, P_i)$,
to be canonical [14], the following basic PBs defined in the transformed
cotangent manifold
$$
\begin{array}{lcl}
\Bigl \{ X_{i}, P_{j} \Bigr \}_{(PB)}^{(x,p)} = \delta_{ij},\; \qquad
\Bigl \{ X_{i}, X_{j} \Bigr \}_{(PB)}^{(x,p)} = 0,\; \qquad
\Bigl \{ P_{i}, P_{j} \Bigr \}_{(PB)}^{(x,p)} = 0, 
\end{array}\eqno (3.15) 
$$
should retain their form. Here the superscripts on the brackets
denote the definition of
the PBs w.r.t. the original phase variables $(x_i, p_i)$. It is 
straightforward to check that the above brackets are not preserved under
the transformations (3.5) and (3.6) if we exploit the
basic PBs defined in (2.3). Thus, the latter transformations are a very
{\it special class} of transformations in the phase space
which are not canonical in nature {\it per se} but they preserve the
sum of the areas in the phase space. \\

\noindent
{\bf 4 Deformed $q$-brackets: connection with quantum groups}\\

\noindent
In this section, we briefly recapitulate some of the key 
and pertinent points of our earlier
works [15,16] on the consistent construction of the dynamics in the
noncommutative phase space where the noncommutativity (i.e. $q$-deformation)
was introduced in the $q$-deformed $2N$ 
dimensional cotangent manifold corresponding to
a given $N$-dimensional configuration manifold. In our present context
(where we are discussing a 2D physical system), the following relationships
among the phase variables $(x_i, p_i)$ (with $i = 1, 2$) 
on the 4D cotangent manifold
$$
\begin{array}{lcl}
x_i\; x_j = x_j \; x_i,\; \qquad p_i\; p_j = p_j \; p_i,\; \qquad
x_i \; p_j = q\; p_j\; x_i,
\end{array}\eqno(4.1)
$$
play a very crucial and decisive role in our whole discussion of 
the dynamics. In fact, the above
$q$-deformation is chosen in the phase space so that the quantum group 
invariance and the ordinary rotational invariance can
be maintained together
\footnote{ This deformation can be easily generalized to the $2N$-dimensional
Minkowski cotangent manifolds for the dynamical discussion of a physical
relativistic system where the quantum group invariance and Lorentz invariance
are respected together for any arbitrary ordering of the spacetime 
indices [15,16].}
{\it for any arbitrary ordering of the indices
$i$ and $j$} (see, e.g., [15] for details). 
We shall concentrate here on the following general 
transformations on the phase variables $(x_i, p_i)$ defined on
the 4D $q$-deformed cotangent manifold 
$$
\begin{array}{lcl}
x_i \rightarrow X_i = A\; x_i + B\; p_i,\; \qquad
p_i \rightarrow P_i = C\; x_i + D\; p_i. 
\end{array}\eqno(4.2)
$$
In the above, the original phase variables $(x_i, p_i)$ 
and the transformed sets of the phase variables $(X_i, P_i)$ 
are assumed to commute with the elements $A, B, C, D$ of a $2 \times 2$ 
matrix corresponding to the quantum group $GL_{q,p} (2)$. Here the deformation
parameters $q, p$ are the non-zero complex numbers 
(i.e. $q, p \in {\cal C} / \{ 0 \}$) corresponding to the most general
quantum group deformation of the ordinary general linear group 
$GL (2)$ of $2 \times 2$ nonsingular matrices. In fact, the elements of the
quantum group $GL_{q,p} (2)$
obey (see, e.g., [17] for details)
$$
\begin{array}{lcl}
&& A B = p B A,\; \quad A C = q C A,\; \quad B C = (q/p) C B,\; \quad
B D = q D B, \nonumber\\
&& C D = p D C,\; \qquad A D - D A = (p - q^{-1}) B C = (q - p^{-1}) C B.
\end{array}\eqno(4.3)
$$
It can be checked that the relationships in (4.1) remain 
{\it form-invariant} under the general transformations (4.2) {\it only
for the choice $ pq = 1$}. In other words, the $q$-deformation (4.1) in the
4D cotangent manifold remains form-invariant under the transformations
generated by the quantum group $GL_{q,q^{-1}} (2)$. It should be noted that
these transformations are different from the 
transformations generated by the single parameter $q$-deformed quantum
group $GL_{q} (2)$. In fact, the latter is the limiting case of 
$GL_{q,p} (2)$ when $ q = p$. 
Under the restriction $ pq = 1$, the relations (4.3) yield the following
braiding relationships among the rows and columns
$$
\begin{array}{lcl}
&& A B = q^{-1} B A,\; \qquad A C = q C A,\; \qquad B C = q^2\; C B, \nonumber\\
&& B D = q D B,\; \qquad
C D = q^{-1}\; D C,\; \qquad A D = D A.
\end{array}\eqno(4.4)
$$
At this stage, it is important 
to note that it is the {\it individual pairs} of the
original phase variables $(x_i, p_i)$ (with $i = 1, 2$) that change to
the transformed phase variables $(X_i, P_i)$ (with $ i = 1, 2$) under
the quantum group $GL_{q, q^{-1}} (2)$. Thus, for our present discussion, 
there are
two sets of quantum groups $GL_{q, q^{-1}} (2)$ which are responsible for
the transformations of the original two pairs of phase variables $(x_1, p_1)$
and $(x_2, p_2)$ in (4.2). This statement can be mathematically expressed, in
the matrix form, as given below
$$
\begin{array}{lcl}
\left (\begin{array} {c} x_{1} \\ p_{1} \\
\end{array} \right )
\;\;\rightarrow \;\;
\left (\begin{array} {c}
X_{1} \\ P_{1} \\
\end{array} \right ) &\;=\;& 
\left (\begin{array} {cc}
A & B\\ C & D\\
\end{array} \right )\;\;
\left (\begin{array} {c}
x_{1} \\ p_{1} \\
\end{array} \right ), \nonumber\\
\left (\begin{array} {c}
x_{2} \\ p_{2} \\
\end{array} \right )
\;\;\rightarrow \;\;
\left (\begin{array} {c}
X_{2} \\ P_{2} \\
\end{array} \right ) &\;=\;& 
\left (\begin{array} {cc}
A & B\\ C & D\\
\end{array} \right )\;\;
\left (\begin{array} {c}
x_{2} \\ p_{2} \\
\end{array} \right ), 
\end{array}\eqno(4.5)
$$
where $A, B, C, D$ are the elements of the quantum group $GL_{q, q^{-1}} (2)$
that satisfy (4.4).

Now the stage is set for establishing the connection between the
noncommutativity discussed in the previous sections and the noncommutativity
due to the quantum groups in the present section. As argued in the paragraph 
after (3.4), one of the key as well as crucial features of the transformations
in (3.5) and (3.6) is the invariance of the sum of the phase space areas. In 
the language of the $GL_{q, q^{-1}} (2)$ invariant differential geometry 
defined on the 4D cotangent manifold (see, e.g., [15] for details), this 
statement can be mathematically captured in the following wedge products and
their transformations
$$
\begin{array}{lcl}
(dx_i \wedge dx_i) &\rightarrow & (d X_i \wedge d X_i) = 0,\; \qquad
(dp_i \wedge dp_i) \rightarrow  (d P_i \wedge d P_i) = 0, \nonumber\\
(d x_i \wedge d p_i) & \rightarrow & (d X_i \wedge d P_i) =
(A D - q^{-1} B C)\; (dx_i \wedge dp_i),
\end{array}\eqno(4.6)
$$
where, in addition to (4.4), we have used the following relationships [15]
$$
\begin{array}{lcl}
(dx_i \wedge dx_i) = 0,\; \qquad
(dp_i \wedge dp_i) = 0,\; \qquad
(d x_i \wedge d p_i) = - q\; (dp_i \wedge dx_i).
\end{array}\eqno(4.7)
$$
This shows that, for the area preserving phase space transformations in (4.6),
we have
$$
\begin{array}{lcl}
A D - q^{-1} B C = A D - q C B = D A - q^{-1} B C = D A - q C B = 1,
\end{array}\eqno(4.8)
$$
which is nothing but setting the $q$-determinant 
(see, e.g., [17] for details) of the $2 \times 2$
$GL_{q, q^{-1}} (2)$ matrix equal to one. In more sophisticated language,
the requirement of the area preserving phase transformations in (4.5) entails
upon the quantum group $GL_{q, q^{-1}} (2)$ to become $SL_{q, q^{-1}} (2)$
as the $q$-determinant $ AD - q^{-1} B C = 1$. From the point of view of the
structure of the PBs defined in the canonical
form (cf. (2.3)) and the nontrivial forms (3.3) and (3.4), it can be seen
that the crucial common feature is
$$
\begin{array}{lcl}
\Bigl \{ x_{i}, p_{j} \Bigr \}_{(PB)}^{(x,p)} = \delta_{ij},\; \qquad
\Bigl \{ x_{i}, P_{j} \Bigr \}_{(PB)}^{(x,p)} = \delta_{ij},\; \qquad
\Bigl \{ X_{i}, p_{j} \Bigr \}_{(PB)}^{(x,p)} = \delta_{ij}. 
\end{array} \eqno(4.9)
$$
In other words, the PBs between the original canonically conjugate
variables $(x_i, p_i)$ and the transformed phase variables $(X_i, p_i)$
and $(x_i, P_j)$ remain unchanged. For the same to remain sacrosanct even 
in the case of
quantum group transformations in (4.5) (or (4.2)), it can be seen that
the following condition
$$
\begin{array}{lcl}
\Bigl \{ X_{i}, P_{j} \Bigr \}_{(PB)}^{(q)} = \Bigl ( A D - q B C \Bigr )\;
\delta_{ij} \; \rightarrow\; 
\Bigl \{ X_{i}, P_{j} \Bigr \}_{(PB)}^{(q)} =  \delta_{ij},
\end{array}\eqno(4.10)
$$
has to be satisfied. In the derivation of the
above, we have exploited the following $q$-deformed
basic Poisson brackets for the phase variables $(x_i, p_i)$ [15]
$$
\begin{array}{lcl}
\Bigl \{ x_{i}, p_{j} \Bigr \}_{(PB)}^{(q)} = \delta_{ij}, \;\quad
\Bigl \{ p_{i}, x_{j} \Bigr \}_{(PB)}^{(q)} = - \;q\; \delta_{ij}, \;\quad
\Bigl \{ p_{i}, p_{j} \Bigr \}_{(PB)}^{(q)} =
\Bigl \{ x_{i}, x_{j} \Bigr \}_{(PB)}^{(q)} = 0,
\end{array} \eqno(4.11)
$$
which crucially depend on the choice of the symplectic structures on the
$q$-deformed 4D cotangent manifold. In our earlier work [15], we have 
chosen the following
$$
\begin{array}{lcl}
\Omega^{AB} (q) =
\left (\begin{array} {cccc}
0 & 0 & 1 & 0 \\
0 & 0 & 0 & 1 \\
- q & 0 & 0 & 0\\
0 & - q & 0 & 0\\
\end{array} \right ),
\qquad \;
\Omega_{AB} (q) =
\left (\begin{array} {cccc}
0 & 0 & - q^{-1} & 0 \\
0 & 0 & 0 & - q^{-1} \\
1 & 0 & 0 & 0\\
0 & 1 & 0 & 0\\
\end{array} \right ).
\end{array}\eqno(4.12)
$$
It is now unequivocally clear that the requirement of the equality 
and consistency between
(4.9) and (4.10) imposes the following restriction 
on the elements of the $GL_{q, q^{-1}} (2)$
$$
\begin{array}{lcl}
A D - q B C = A D - q^3 C B = D A - q B C = D A - q^3 C B = 1.
\end{array}\eqno(4.13)
$$
This shows that, for a theory in which the canonical brackets (4.9)
and the area preserving transformations (i.e., $ d X_i \wedge d P_i
= d x_i \wedge d p_i$) are respected together, we obtain a restriction
$ q^2 = 1$ on the deformation parameter. This condition emerges from 
the equality of (4.8) and (4.13).
With the $q$-brackets defined in (4.11), it is straightforward
to check (using the $q$-commutation relations (4.4)) that the following
brackets, defined in the 2D coordinate as well as the momentum plane
of the 4D q-deformed cotangent manifold, are true; namely,
$$
\begin{array}{lcl}
\Bigl \{ X_{i}, X_{j} \Bigr \}_{(PB)}^{(q)} &=& 
(AB - q BA)\; \delta_{ij}
\equiv (1 - q^2)\; A B\; \delta_{ij}, \nonumber\\
\Bigl \{ P_{i}, P_{j} \Bigr \}_{(PB)}^{(q)} &=&
(C D - q D C)\; \delta_{ij}
\equiv (1 - q^2)\;C D\; \delta_{ij}.
\end{array}\eqno(4.14)
$$
These $q$-brackets turn out to be zero for $q^2 = 1$.
Thus, it is clear that the NC structure 
in the 2D coordinate as well as momentum planes (cf. (3.3) and (3.4))
does {\it not} exist for
our quantum group considerations if we require the consistency between the
$q$-deformed PBs and the area preserving 
$SL_{q, q^{-1}} (2)$ invariant phase space transformations. Hence, it is
obvious that for $q^2 = 1$, the quantum group considerations allow the phase
space transformations that are exactly canonical in nature.\\

\noindent
{\bf 5 Conclusions}\\

\noindent
In the present investigation, we have established (i) a connection between
the canonical PBs and the nontrivial NC brackets, and (ii)
the relation between the canonical Hamiltonian and the nontrivial Hamiltonians.
These connections emerge because of the existence of a
special class of symmetry transformations
(cf. (3.5) and (3.6)) on the phase variables that
preserves the sum of the areas on the (un-)transformed cotangent manifolds.
This symmetry transformation is, however, shown {\it not}
to be exactly canonical in nature unlike the canonical transformations of
(3.8) which leave the canonical brackets form-invariant.
As far as the dynamics on the 4D tangent manifold is concerned, it remains
unaffected by the NC structures present in the PBs {\it equivalently} defined
on the 2D coordinate or momentum planes that are
embedded in the 4D cotangent manifolds. This fact is captured in the
appearance of the second-order Lagrangian $L^{(s)} (x, \dot x)$ which remains
unchanged irrespective of the noncommutativity
present in the PBs defined on the 4D cotangent manifold.
The above cited ``sum of the area preserving transformation'' is also
reflected in the realm of quantum group 
transformations on the phase variables where the transformations,
generated by $GL_{q, q^{-1}} (2)$, are restricted to become the
transformations corresponding to the quantum group $SL_{q, q^{-1}} (2)$.
The requirement of the consistency of the above transformations with
the structure of the $q$-deformed PBs, however, puts a restriction 
$q^2 = 1$ on the deformation parameter. It is interesting to
point out that, in a completely different context, exactly the same
restriction (i.e. $q = \pm 1$)) has been shown to exist for 
the requirement of equivalence between the gauge- and reparametrization 
invariances for a $q$-deformed relativistic (super-)particle moving
in an undeformed $D$-dimensional target space [16].

\end{document}